\documentclass[12pt]{article}
\usepackage{amssymb,amsmath}
\usepackage[noblocks]{authblk}
\usepackage[top=0.75in, bottom=0.75in, left=0.75in, right=0.75in, dvips]{geometry}
\usepackage{caption}
\begin{document}
\textwidth 10.0in 
\textheight 9.0in 
\topmargin -0.60in
\title{Supersymmetry and the Rotation Group}
\author[1,2]{D.G.C. McKeon}
\affil[1] {Department of Applied Mathematics, The
University of Western Ontario, London, ON N6A 5B7, Canada} 
\affil[2] {Department of Mathematics and
Computer Science, Algoma University, \newline Sault Ste. Marie, ON P6A
2G4, Canada}
\date{}
\maketitle          

\maketitle      
\noindent
email: dgmckeo2@uwo.ca

\begin{abstract}
A model invariant under a supersymmetric extension of the rotation group $0(3)$ is mapped, using a stereographic projection, from the spherical surface $S_2$ to two dimensional Euclidean space.  The resulting model is not translation invariant.  This has the consequence that fields that are supersymmetric partners no longer have a degenerate mass. This degeneracy is restored once the radius of $S_2$ goes to infinity, and the resulting supersymmetry transformation for the fields is now mass dependent.  An analogous model on the surface $S_4$ is introduced, and its projection onto four dimensional Euclidean space is examined.  This model in turn suggests a supersymmetric model on $(3+1)$ dimensional Minkowski space.
\end{abstract}

\section{Introduction}
The relationship between the conformal group [1,2] in Minkowski space and the anti-de Sitter group is well established [3,4].  An analogue exists between the conformal group in Euclidean space and the rotation group.  This has led to a mapping of models in an $n + 1$ dimensional spherical space to an $n$ dimensional Euclidean space [5,6].

It has been possible to construct models that are invariant under supersymmetric extensions of the invariance present in spaces of constant curvature (such as spherical or anti-de Sitter spaces) [7-11].  In such models the generators of the supersymmetry transformation are no longer the ``square root'' of the generator of translations. As a result there is no degeneracy between the Boson and Fermion masses as occurs in the supersymmetric extension of the Poincar$\acute{\rm{e}}$ group. This degeneracy is a major problem when trying to construct supersymmetric models that are phenomenologically viable.

In this paper we seek to avoid the degeneracy in masses between Bosonic and Fermionic fields that are supersymmetric partners. To show how this might be done, we first examine a model on the sphere $S_2$ of radius $R$ that is invariant under a supersymmetric extension of the group $0(3)$ and show how a stereographic projection can be used to project it onto two dimensional Euclidean space.  The resulting model is explicitly dependent on $x^2$ so that it is not translationally invariant. This lack of translation invariance means that it becomes possible to break the degeneracy in Bosonic and Fermionic masses.  As $R \rightarrow \infty$, this degeneracy as well as translational symmetry are restored. We then use this model to introduce a supersymmetric model on the sphere $S_4$ and consider its projection onto four dimensional Euclidean space, again taking the limit $R \rightarrow \infty$.  A version of this model in $(3 + 1)$ dimensional Minkowski space is given.

\section{The Model}

In three Euclidean dimensions, irreducible spinors are two component Dirac spinors. If the generator of rotations is $J^a$, of supersymmetry transformations is $Q_i$ and $Z$ is a ``central charge'', then a suitable supersymmetry algebra is [10,11]
\begin{subequations}
\begin{align}
\left[ J^a, J^b \right] &= i\epsilon^{abc}J^c\\
\left[ J^a, Q_i \right] &= -\frac{1}{2} \tau^a_{ij} Q_j\\
\left[ Z, Q_i \right] &= \mp Q_i\\
\left\lbrace Q_i, Q_j^\dagger \right\rbrace &= Z\delta_{ij} \mp 2 \tau_{ij}^a J^a
\end{align}
\end{subequations}
where $\tau^a$ is a Pauli spin matrix satisfying
\begin{equation}
\tau^a \tau^b = \delta^{ab} + i\epsilon^{abc}\tau^c.
\end{equation}
Representations of this algebra are discussed in ref. [10]; it is shown there that the eigenvalues of $Z$ form an upper bound to the eigenvalues of $J^2$.

A supersymmetric action for a complex scalar $\Phi$, a two component Dirac spinor $\Psi$ and an auxiliary complex scalar $F$ is [10]
\begin{align}
S_2 = \int \frac{dA}{R^2} & \Big[ \Big( \frac{1}{2} \Psi^\dagger ( \tau \cdot L + \rho) \Psi - \Phi^*\left(L^2 + \rho(1-\rho)\right)\Phi\nonumber \\
& - \frac{1}{4} F^*F\Big) + \lambda_N \Big( 2(1-2\rho) \Phi^*\Phi - (F^*\Phi + F\Phi^*)\nonumber \\
& \hspace{1cm} - \Psi^\dagger\Psi\Big)^N\Big] \qquad \left(L^a \equiv -i\epsilon^{abc} \eta^b \partial^c , \quad \rho = {\rm{constant}}\right).
\end{align}
In eq. (3), $N$ is a positive integer and $\lambda_N$ a coupling constant. 
This action is invariant under the supersymmetry transformation
%4,5
\begin{subequations}
\begin{align}
\delta \Phi &= \xi^\dagger\Psi\\
\delta \Psi &= 2(\tau \cdot L + 1 - \rho) \Phi \xi - F\xi\\
\delta F &= -2\xi^\dagger  (\tau \cdot L + \rho) \Psi
\end{align}
\end{subequations}
as well as
\begin{subequations}
\begin{align}
\delta \Phi &= \sigma i \left[ 2(1-\rho)\Phi - F\right]\\
\delta \Psi &= \sigma i (1 +2 \tau \cdot L) \Psi \\
\delta F &= \sigma i \left[ -4\left( L^2 + \rho(1-\rho)\right) \Phi + 2\rho F\right]
\end{align}
\end{subequations}
where $\xi$ is a Grassmann two component Dirac spinor and $\sigma$ is a real scalar (both are constants).  A surface element on the sphere $\eta^2 = R^2$ embedded in three dimensions is $dA$.  The transformations of eqs. (4,5) are consistent with the algebra of eq. (1). This is in part because the operator $L^a$ in eq. (3) satisfies the same algebra as the generator of rotations $J^a$ (eq. (1a)).

If now we define
%6
\begin{equation}
\Lambda_\pm = \frac{1\pm i\tau\cdot\eta/R}{\sqrt{2}} = (\Lambda_{\mp})^\dagger = (\Lambda_{\mp})^{-1}
\end{equation}
then as by eq. (2) 
%7
\begin{equation}
(\tau \cdot L + 1) \tau \cdot \eta = -\tau \cdot \eta (\tau \cdot L + 1)
\end{equation}
we have
%8
\begin{equation}
\Lambda_\pm (\tau \cdot L + 1) \Lambda_\mp = \pm i \frac{\tau \cdot \eta}{R}(\tau \cdot L + 1).
\end{equation}

By eq. (8), if $\Psi = \Lambda_-\Psi^\prime$, $\xi = \Lambda_-\xi^\prime$, then the Fermionic terms in the model of eq. (3) become
%9
\begin{align}
S_\Psi = \int \frac{dA}{R^2}\Bigg[ \frac{1}{2} \Psi^{\prime\dagger} & \left( \frac{i\tau \cdot \eta}{R}(\tau \cdot L+1) - (1-\rho)\right)\Psi^\prime \nonumber \\
& + \lambda_N \left[- \Psi^{\prime\dagger} \Psi^\prime\right]^N\Bigg].
\end{align}
The invariances of eqs. (4,5) are similarly transformed.

We now consider how a stereographic projection can be used to map a sphere in $(n + 1)$ dimensions onto an $n$ dimensional plane. In refs. [5,6] it is shown that under the change of variable from $x^a$ to $\eta^a$ in $(n+1)$ dimensions
%10
\begin{equation}
\eta^a = R \mathit{h}^a + 2 \frac{x^a-h^ax^2/R}{1-2x\cdot h/R + x^2/R^2}
\end{equation}
the plane $x \cdot h = 0$ is mapped onto the sphere $\eta^2 = R^2$.  (The vector $h^a$ is a unit vector in the direction of $\eta^{n+1}$.)  On this plane we have the stereographic projection $\eta^{n+1} = R\left(\frac{1-x^2/R^2}{1+x^2/R^2}\right)$ and $\eta^i = \frac{2x^i}{1+x^2/R^2}(i = 1 \ldots n)$. It follows that if
%11
\begin{equation}
\kappa = 1 + \eta^{n+1}/R = \frac{2}{1+x^2/R^2}
\end{equation}
then [5,6]
%12
\begin{equation}
d^nx = \kappa^{-n} dA.
\end{equation}
The fields $\Phi$ and $\Psi^\prime$ map onto fields $\phi$ and $\psi$ so that [12,13] in general for $n$ dimensions
%13
\begin{subequations}
\begin{align}
\Phi &= \kappa^{1-n/2}\phi\\
\Psi^\prime &= \kappa^{(1-n)/2} (\kappa/2)^{1/2}(1 - \mathit{h}\cdot \alpha\; x \cdot \alpha/R)\psi\\
&\equiv \kappa^{(1-n)/2} U\psi\quad (U^{-1} = U^\dagger) \nonumber
\end{align}
\end{subequations}
where $\alpha^a$ is a matrix in $n + 1$ dimensional space satisfying the Clifford algebra
%14
\begin{equation}
\left\lbrace \alpha^a, \alpha^b\right\rbrace = 2\delta^{ab}.
\end{equation}
If now
%15
\begin{equation}
\gamma^{ab} = \frac{-1}{4}\left[ \alpha^a, \alpha^b \right]
\end{equation}
and
%16
\begin{equation}
L^{ab} = -\left(\eta^a \partial^b - \eta^b\partial^a\right)
\end{equation}
then in three dimensions where $\alpha^a = \tau^a$ (by eq. (2)), $\gamma^{ab} = \frac{-1}{2} i\epsilon^{abc}\tau^c$ and $L^{ab} = -i\epsilon^{abc}L^c$ with $L^c$ defined in eq. (3) so that
%17
\begin{subequations}
\begin{align}
\tau \cdot L &= - \gamma^{ab} L^{ab}\\
L^{ab} L^{ab} &= -L^aL^a.
\end{align}
\end{subequations}
It can then be shown that [12] under the transformation of eq. (13), in $n$ dimensions
%18
\begin{equation}
\int dA \Phi^* \left( \frac{L^{ab}L^{ab} - \frac{1}{2}n(n-2)}{2R^2} \right)\Phi = \int d^nx \;\phi^* \partial^2 \phi
\end{equation}
and [13]
%19
\begin{equation}
\int dA \Psi^{\prime\dagger} \left( \frac{\alpha \cdot \eta (\gamma^{ab}L^{ab} - n/2)}{R^2} \right)\Psi^\prime = \int d^n \;x \psi^\dagger \alpha \cdot \partial \psi.
\end{equation}
We now can revert to $n = 2$ dimensions. Eqs. (12,13,17,18,19) show that the model of eqs. (3,9) becomes in two dimensional Euclidean space 
\begin{align}
S_2 = \int d^2x \Bigg[ \Bigg( - \frac{1}{2} & \psi^\dagger \left( \frac{i}{R}\tau \cdot \partial + \frac{\kappa(1-\rho)}{R^2}\right)\psi + \phi^*\left( \partial^2 - \frac{\kappa^2 \rho(1-\rho)}{R^2}\right)\phi\nonumber \\
& - \frac{\kappa^2}{4R^2} f^*f\Bigg) + \frac{\lambda_N\kappa^2}{R^2}\Bigg(2 (1-2\rho) \phi^*\phi - (f^*\phi + f\phi^*) - \kappa^{-1}\psi^\dagger\psi\Bigg)^N\Bigg]
\end{align}
where $F = f$ (from eq. (13a)).  Since $RU \tau \cdot hU^{-1} = \tau \cdot \eta$, it follows from eq. (13) that eq. (4) becomes
%21
\begin{subequations}
\begin{align}
\delta\phi &= \kappa^{-1} \zeta^\dagger\psi\\
\delta f &= 2i\kappa^{-2} R \zeta^\dagger \tau_i \frac{\partial}{\partial x^i} \psi + 2\kappa^{-1} (1-\rho) \zeta^\dagger\psi\\
\delta \psi &= -2\left( i\kappa^{-1} R\tau_i\frac{\partial}{\partial x_i} + \rho\right) \phi\zeta - f\zeta .
\end{align}
\end{subequations}
Here $\xi = \kappa^{-1/2} U\zeta$ as in eq. (13b).  Both the model of eq. (20) and the symmetry of eq. (21) lack translation invariance because of the contribution of $\kappa$ defined in eq. (11). However, because the supersymmetry transformation is no longer a ``square root`` of a translation, the fields $\phi$ and $\psi$ in eq. (20) no longer have degenerate masses.

Upon making the rescalings
%22
\begin{equation}
\psi \rightarrow \sqrt{R}\psi, \quad \rho \rightarrow R\rho, \quad 
\phi \rightarrow \phi, \quad f \rightarrow Rf, \quad \zeta \rightarrow \zeta/\sqrt{R}, \quad \lambda_N \rightarrow \frac{\lambda_N R^{2-N}}{4}
\end{equation}
and letting $R \rightarrow \infty$, eqs. (20) and (21) become
%23
\begin{align}
S_{2\infty} &= \int d^2x \Big[ - \frac{1}{2} \psi^\dagger \left(i \tau \cdot \partial - 2\rho\right) \psi + \phi^* \left( \partial^2 - 4 \rho^2\right)\phi - f^*f\\
&\qquad + \lambda_N \left( -4 \rho \phi^*\phi - \left(f \phi^\dagger + f^*\phi\right) - \frac{1}{2} \psi^\dagger \psi \right)^N \Big]\nonumber
\end{align}
and
%24
\begin{align}
\delta\phi &= \frac{1}{2} \zeta^\dagger\psi \\
\delta f &= \frac{i}{2}\zeta^\dagger \tau \cdot \partial \psi - \rho \zeta^\dagger \psi  \nonumber \\
\delta \psi &= - \left( i \tau  \cdot \partial + 2\rho\right) \phi \zeta - f\zeta \nonumber
\end{align}
respectively.  Similarly, as $R \rightarrow \infty$, the transformations of eq. (5) become
%25
\begin{align}
\delta \phi &= -\sigma i (2 \rho \phi + f)\\
\delta \psi &= 4\sigma \tau \cdot \partial \psi\nonumber \\
\delta f &= 2\sigma i \left[ 2 \left( \partial^2 + \rho^2\right)\phi + \rho f\right].\nonumber
\end{align}

Upon letting $R \rightarrow \infty$ all dependence of the action $S_{2\infty}$ and of the supersymmetry transformation on $x^2$ is lost and translational symmetry is restored.  We note that after eliminating $f$ and $f^*$ from the action by using their equations of motion when $\lambda_N = 0 (N > 1)$, the terms in $S_\infty$ that are bilinear in $\psi^\dagger\psi$ and $\phi^*\phi$ are respectively $\left(\rho - \frac{\lambda_1}{2}\right)\psi^\dagger\psi$ and 
$4\left(\rho - \frac{\lambda_1}{2}\right)^2\phi^*\phi$ and so the degeneracy between the Boson and Fermion masses is restored.

The supersymmetry transformation of eq. (24) is unlike the one occurring in the original Wess-Zumino model [14] in that it involves the dimensionful parameter $\rho$, and the symmetry of eq. (25) is novel. These two symmetries are a result of the supersymmetry algebra of eq. (1).

Two supersymmetric extensions of the $0(5)$ algebra are [10]
%26
\begin{subequations}
\begin{align}
\left[ J^{AB}, Q_i \right] &= - \Sigma_{ij}^{AB} Q_j\\
\left[ Z, Q_i \right] &= \mp Q_i\\
\left\lbrace Q_i, Q_j^\dagger \right\rbrace &= Z \delta_{ij} \pm \Sigma_{ij}^{AB} J^{AB}
\end{align}
\end{subequations}

and
%27
\begin{subequations}
\begin{align}
\left[ J^{AB}, Q_i \right] &= - \Sigma_{ij}^{AB} Q_j\\
\left[   J^{AB}, Z^C\right] &= \delta^{AC}Z^B - \delta^{BC} Z^A\\
\left[ Z^A, Q_i \right] &= -\frac{1}{2} \gamma_{ij}^A Q_j\\
\left[ Z, Q_i \right] &= -Q_i\\
\left[ Z^A, Z^B \right] &= - J^{AB}\\
\left\lbrace Q_i, Q_u^\dagger \right\rbrace &= \pm \left( \frac{3}{2} Z\delta_{ij} - \gamma_{ij}^A Z^A + \Sigma_{ij}^{AB} J^{AB} \right).
\end{align}
\end{subequations}
The fact that these algebras are related to the supersymmetry algebra of eq. (1) suggests that the following action on a sphere $S_4$ of radius $R$
%28
\begin{align}
S_4 = \int \frac{dA}{R^4} &\Big\{ \Big[ \Psi^\dagger \left( \Sigma^{AB} L^{AB} + \mu\right) \Psi + \frac{1}{2} \Phi^* \left(L^{AB}L^{AB} + 2\mu (\mu + 3)\right)\Phi\nonumber \\
&- F^*F \Big] + \lambda_N \Big[ (2\mu +3) \Phi^*\Phi - (F^*\Phi + F\Phi^*)\nonumber \\
&\hspace{1cm} - \Psi^\dagger \Psi\Big]^N\Big\} \quad (N = 1,2,3 \ldots)
\end{align}
where $\Psi$ is a four component Dirac spinor, $\Phi$ and $F$ are complex scalers, and $\mu$ and $\lambda_N$ real constants.

Using the identity
%29
\begin{subequations}
\begin{align}
\gamma^A\gamma^B\gamma^C &= \delta^{AB} \gamma^C - \delta^{AC} \gamma^B + \delta^{BC} \gamma^A + \epsilon^{ABCDE} \Sigma^{DE}
\intertext{it follows that}
\left(\Sigma^{AB} L^{AB}\right)^2 &= -\frac{1}{2} (L^{AB})^2 + 3 \Sigma^{AB}L^{AB}\\
\left(\Sigma^{AB} L^{AB}-2\right)\gamma^C\eta^C  &= -\gamma^C\eta^C\left(\Sigma^{AB}L^{AB} - 2\right).
\end{align}
\end{subequations}
From this, it is possible to show that eq. (28) is invariant under the supersymmetry transformation
%30
\begin{subequations}
\begin{align}
\delta \Psi &= \left( \Sigma^{AB} L^{AB} - \mu - 3\right) \Phi \xi + F \xi\\
\delta \Phi &=  \xi^\dagger  \Psi\\
\delta F &=  \xi^\dagger \left( \Sigma^{AB} L^{AB} + \mu\right) \Psi 
\end{align}
\end{subequations}
where $\xi$ is a constant Grassman spinor.

Again using the stereographic projection of eqs. (10, 13) and then after rescaling
%31
\begin{equation}
\psi \rightarrow R^{3/2}\psi, \quad \mu \rightarrow R\mu, \quad \phi \rightarrow R\phi, \quad f \rightarrow R^2 f,\quad \lambda_N \rightarrow R^{4-3N}\lambda_N/16
\end{equation}
and letting $R \rightarrow \infty$, we end up with the action 
%32
\begin{align}
S_{4\infty} = \int d^4x \Bigg\{ & \psi^\dagger \left( i\gamma \cdot \partial + 2\mu \right) \psi + \phi^* \left( \partial^2 + 4\mu^2\right)\phi - 4 f^*f\nonumber \\
&+ \lambda_N \left[ \frac{1}{2}\mu \phi^*\phi - \frac{1}{4} (f^*\phi + f\phi^*) + \frac{1}{8} \psi^\dagger \psi\right]^N\Bigg\}
\end{align}
which is invariant under the supersymmetry transformation 
%33
\begin{subequations}
\begin{align}
\delta \psi &= \left( \frac{i}{4} \gamma \cdot \partial - \frac{\mu}{2}\right)\phi \zeta + \frac{1}{2} f\zeta\\
\delta \phi &= \frac{1}{4}\zeta^\dagger \psi\\
\delta f &= \zeta^\dagger \left( \frac{i}{8} \gamma \cdot \partial + \frac{\mu}{4}\right)\psi.
\end{align}
\end{subequations}
Eqs. (32, 33) are four dimensional versions of eqs. (23, 24).  They also suggest a supersymmetric model in $(3 + 1)$ dimensional Minkowski space; it is 
%34
\begin{align}
S_M = \int d^4 x & \Bigg\{ \overline{\psi} (i \gamma \cdot \partial + 2 \mu) \psi + \phi (\partial^2 + 4\mu^2) \phi - 4f^2\\
&+ \lambda_N \left[ \frac{1}{2} \mu\phi^2 - \frac{1}{2} f\phi + \frac{1}{8} \overline{\psi} \psi \right]^N\Bigg\}\nonumber
\end{align}
where now $\psi$ is a Majorana spinor and $f$ and $\phi$ real scalars. The action $S_M$ is invariant under the supersymmetry transformation
%35
\begin{subequations}
\begin{align}
\delta \psi &= \left( \frac{i}{4} \gamma \cdot \partial - \frac{\mu}{2}\right) \phi \zeta + \frac{1}{2} f \zeta\\
\delta \phi &=  \frac{1}{4} \overline{\zeta} \psi\\
\delta f &= \overline{\zeta} \left( \frac{i}{8} \gamma \cdot \partial + \frac{\mu}{4}\right)\psi.
\end{align}
\end{subequations}
If $\lambda_N = 0(N > 1)$ in eq. (32), the equation of motion for $f$ can be used to eliminate $f$ leaving us with
%36
\begin{equation}
S_M = \int d^4x \left[ \overline{\psi}\left( i \gamma \cdot \partial + \left( 2\mu + \frac{\lambda_1}{8}\right)\right)\psi + \phi\left(\partial^2 + \left(2\mu +\frac{\lambda_1}{8}\right)^2\right)\phi\right].
\end{equation}

\section{Discussion}

The model of eq. (20) is unusual, having explicit dependence on $x^2/R^2$.  However, it does possess the supersymmetry of eq. (21) as a consequence of the original model of eq. (3) being invariant under the transformations of eqs. (4,5).  Since this supersymmetry is not the ``square root'' of a generator of translations, there is no degeneracy between the Boson and Fermion masses. This degeneracy is restored if $R \rightarrow \infty$. Similar models have been devised in $(4 + 0)$ and $(3 + 1)$ dimensions.

We are examining if a more realistic model which incorporates supersymmetry in this fashion can be devised. In particular, it is hoped that a model in $4 + 2$ dimensions with a supersymmetry that is an extension of an $S0(4,2)$ symmetry can be found and that this model can be projected onto four dimensional Minkowski space.  Radiative corrections to the models of eqs. (3, 28) are also being considered.

\section*{Acknowledgements}
This work is a result of a long collaboration with T.N. Sherry. R. Macleod was helpful at an early stage.

\end{document}